\documentclass[aps,pre,reprint, showpacs,amsmath,amssymb,longbibliography]{revtex4-2}
\usepackage{xcolor,hyperref}
\hypersetup{colorlinks,linkcolor={blue!50!black},
urlcolor={blue!80!black}}
\usepackage{graphicx,float}
\usepackage{xcolor}
\usepackage{tabularx}
\usepackage{booktabs}
\usepackage{multirow}

\newcommand{\be}{\begin{equation}}
\newcommand{\ee}{\end{equation}}
\newcommand{\fig}[1]{Fig.~\ref{#1}}
\newcommand{\figs}[1]{Figs.~\ref{#1}}
\newcommand{\Fig}[1]{Figure~\ref{#1}}
\newcommand{\Figs}[1]{Figures~\ref{#1}}

\newcommand{\ta}{\tau_\alpha}

\begin{document}
        \title{Unified percolation scenario for the $\alpha$ and $\beta$ processes in simple glass formers} 
        \author{Liang Gao, Hai-Bin Yu}\email{haibinyu@hust.edu.cn}
        \affiliation{Wuhan National High Magnetic Field Center and School of Physics, Huazhong University of Science and Technology, Wuhan 430074, Hubei, China}
	\date{\today}
	\author{Thomas B. Schr{\o}der, Jeppe C. Dyre}\email{dyre@ruc.dk}
	\affiliation{\textit{Glass and Time}, IMFUFA, Department of Science and Environment, Roskilde University, 
            P.O. Box 260, DK-4000 Roskilde, Denmark}

\begin{abstract}
Given the vast differences in interaction details, describing the dynamics of structurally disordered materials in a unified theoretical framework presents a fundamental challenge to condensed-matter physics and materials science. Here, we numerically investigate a double-percolation scenario for the two most important relaxation processes of supercooled liquids and glasses, the so-called $\alpha$ and $\beta$ relaxations. For several simple glass formers, we find that when monitoring the dynamic shear modulus as temperature is lowered from the liquid state, percolation of immobile particles takes place at the temperature locating the $\alpha$ process. Mirroring this, upon continued cooling into the glass state, the mobile-particle percolation transition pinpoints a $\beta$ process whenever the latter is well separated from the main ($\alpha$) process. For two-dimensional systems under the same conditions, percolation of mobile and immobile particles occurs nearly simultaneously and no $\beta$ relaxation can be identified. Our findings suggest that a general description of glassy dynamics should be based on a percolation perspective.
\end{abstract}

\maketitle

\section{Introduction}

A liquid close to the glass transition relaxes extremely slowly toward equilibrium when subjected to an external disturbance  \cite{har76,ang95,edi96,deb01,dyr06,ber11,hun12,wan12,mck17,alb22}. Depending on the temperature the main so-called $\alpha$ relaxation time, $\ta$, can be seconds, hours, even months, with no other limit than the patience of the experimentalist \cite{ang95,edi96,deb01,dyr06,hec08,mck17,alb22}. In the Maxwell model of viscoelasticity \cite{lam78,dyr06}, a glass-forming liquid behaves like a solid on time scales shorter than $\ta$ and flows on longer time scales. Interestingly, most liquids exhibit additional, faster relaxations. The most prominent one is the Johari-Goldstein $\beta$ process, which is observed in virtually all glass-forming organic liquids, polymers, metallic glasses, etc \cite{mccrum67,dyr03,kremer_broadband,yu14,ric15,luo17}. 

It was known already in the 1960s that polymers exhibit relaxation processes above the $\alpha$ relaxation frequency $1/\ta$, which were attributed to side-chain motion \cite{mccrum67}. It was a great surprise, however, when Johari and Goldstein reported in 1970 that fast processes also occur in glasses of small rigid molecules \cite{joh70}. To explain this they proposed the existence of ``islands of mobility'' \cite{gol69,joh70}; this constituted an early example of the dynamic heterogeneities subsequently identified as a universal feature of glass-forming liquids \cite{edi00,dynhet,ber11,kar14}. Nowadays, the term ``$\beta$ relaxation'' is used for the first relaxation process at frequencies higher than those of the main ($\alpha$) process, independent of its origin. A further relaxation process worth mentioning is the $\beta'$ (or $\gamma$) process, which has been observed in both experiments and molecular dynamics simulations. It originates from high-frequency particle diffusion in glass-forming liquids and is considered an extension of the dynamics observed in high-temperature liquids \cite{cha22, yu24nm}. 

Research in the past decades has demonstrated that $\beta$ relaxation plays a crucial role for the mechanical and thermal properties of amorphous materials \cite{YangNanoLett2021,Zhou2024PMS,Qiao2019PMS,wan19,PengSA2020}.
Ngai has suggested that the $\beta$ process is a precursor of the main $\alpha$ relaxation: before the onset of $\beta$ relaxation one finds a regime in which molecules are confined to cages defined by the anharmonic intermolecular potential \cite{nga05,nga12,nga23}. A related proposal was discussed in 1999 by Kudlik \textit{et al.}, who suggested that the $\beta$ process in molecular liquids is a local, spatially restricted reorientation process preceding the $\alpha$ relaxation \cite{kud99}, an idea that is reminiscent of the fundamental prediction of mode-coupling theory \cite{got08}. Experiments have confirmed the caged-molecule picture of the $\beta$ process by detecting small-angle jumps \cite{kau90}, but interestingly large-angle jumps are sometimes also involved in $\beta$ relaxation \cite{arb96}. 

Although most papers on the $\beta$ process report results for the glass phase, $\beta$ relaxation is present also in the equilibrium (metastable) liquid phase above the glass-transition temperature $T_g$. In most cases, however, the $\beta$ process is here partly merged with the $\alpha$ process and observed only as an wing of the latter \cite{ols98a,sch00b,sca21,gui22}. Lengthy annealing of a glass to approach the liquid phase may in some cases separate the $\beta$ relaxation wing from the $\alpha$ peak and establish it as an independent, well-defined process \cite{sch00b}, but in other cases annealing \textit{annihilates} the $\beta$ process by decreasing its magnitude to below the resolution limit \cite{dyr03}. 

Because a glass-forming liquid is disordered, it must be expected that the energy barriers of flow events vary in space. The existence of a wide barrier distribution was recently documented in simulations by Pica Ciamarra and co-workers \cite{cia24}. Assuming the barriers vary randomly in space amounts to replacing complexity by randomness, which is an old and venerated strategy of simplification \cite{wol92}. Once spatial randomness is introduced into the modeling, the phenomenon of percolation comes to mind by connecting \textit{randomness} and \textit{geometry} \cite{bro57,stauffer_percolation,grimmett,isi92}: If finite-size domains in space are marked randomly one after the other, at some point the marked domains will percolate throughout the sample. The value of the percolation threshold depends on the spatial dimension and on the model in question. In one dimension the percolation threshold is unity. On a 2D cubic lattice the link-percolation threshold is 50\% by self-duality \cite{stauffer_percolation,grimmett,isi92}. The percolation threshold decreases with increasing spatial dimension $D$, and for $D\to\infty$ it approaches zero. The threshold is 0.25 for link percolation on a 3D simple cubic lattice and 0.31 for site percolation on the same lattice \cite{stauffer_percolation,grimmett,isi92}.

Recently a double-percolation picture was proposed linking the $\alpha$ and $\beta$ processes in the liquid phase above $T_g$ to the percolation of immobile and mobile regions, respectively \cite{dyr24}. Associating a given energy barrier, $\Delta E$, with a relaxation time proportional to $\exp(\Delta E/k_BT)$ where $T$ is the temperature, the von Neumann type ``minimax'' idea is that $\tau_\alpha$ is controlled by the lowest energy barriers on the percolation cluster formed by the most immobile regions, i.e., those of largest barriers. On time scales longer than $\tau_\alpha$ this cluster breaks up and flow becomes possible. As a consequence, the solid-like structure maintaining the energy barriers throughout the sample by keeping surrounding molecules in place disappears on time scales longer than $\ta$ \cite{sti88,glo00,pas11,rut12,dou22}. This means that the largest energy barriers are never overcome, hence their light color in \fig{fig:alfabeta}(a) illustrating double percolation. Borrowing a term from NMR \cite{die97,sil99} this phenomenon was long ago referred to as ``exchange'', while nowadays the term ``facilitation'' is preferred \cite{fre84,gar02,rit03,cha21,zha22a,oza23,cos24}. Considering next the opposite limit of very short times, only the lowest-barrier regions are relevant and these are spatially separated. Extended fast motion becomes possible when these regions percolate on longer time scales. This gives rise to a new relaxation channel that we propose is the $\beta$ process \cite{ste10,gao23,dyr24}.

The idea that percolation is important for understanding relaxations in viscous liquids and glasses has a long history \cite{don98,glo00,rus00,lon01,shi05,ste10,pas11,sta13,cic14,yu17,bet18,cap19,cap21,cha22,dou22,spi22,gao23,dyr24}. A double-percolation picture was proposed in 1996 by Novikov \textit{et al.} in the context of percolation of liquid-like and solid-like domains defined by the largest and smallest vibrational mean-square displacement, respectively \cite{nov96}, but no relation to the $\alpha$ and $\beta$ processes was proposed at the time. 

\Fig{fig:alfabeta} illustrates our main idea and how it is tested in simulations. With current computers it is not possible to test numerically the double-percolation scenario in the equilibrium liquid phase in the way this scenario was discussed in Ref. \onlinecite{dyr24}; it would require extremely long low-temperature simulations to separate the $\alpha$ and $\beta$ processes, which as mentioned is a challenge even in experiments lasting a long time. As an alternative, we proceed as follows. \Fig{fig:alfabeta}(a) presents the double-percolation scenario for a linear-response loss at the fixed (angular) frequency $\omega$, $\chi''(\omega)$, monitored as a function of temperature when the system is cooled from the equilibrium liquid state into the glass phase. This procedure allows one to monitor the activation energy distribution in much the same way as a constant-temperature equilibrium-liquid frequency scan. The largest activation energies are probed at high temperatures, with gradually smaller ones becoming relevant as the system is cooled. Two ``percolation temperatures'' are conjectured to pinpoint the $\alpha$ and $\beta$ processes, respectively. At any given time, all particles are marked as either mobile or immobile. The immobile particles percolate below the upper percolation temperature and the mobile particles percolate above the lower percolation temperature. Thus both species percolate between these two temperatures. Since this is not possible in 2D, we predict that separate $\alpha$ and $\beta$ relaxations should not be observed here. Note also that the system is assumed to have an activation-energy distribution much wider than $k_BT$, an assumption that may not always be realistic.

Our simulations follow the strategy of Ref. \onlinecite{gao23}, which mimicked experimental dynamic mechanical spectroscopy (DMS) by occasionally during a slow cooling through the glass transition subjecting the sample to a periodic deformation in order to probe the dynamic shear modulus. The numerical study of $\alpha$ and $\beta$ processes in realistic molecular models is extremely challenging because it requires both specially designed models and lengthy computations \cite{yu17,gui22,shi23,Fragiadakis2012PRE}. For this reason, the present paper tests the double-percolation scenario by simulations of point-particle models.

We find that whenever the $\alpha$ and $\beta$ processes are well separated, they correspond to the percolation transition of immobile and mobile particles, respectively. This confirms the double-percolation scenario. There are also cases where the two processes are not well separated, however; in particular this is always the situation in 2D. Most of the models studied are binary metallic glass formers. Our investigation includes also data on a binary Lennard-Jones system, the metalloid NiP,  and a ternary metallic glass, but relaxations of ionic or covalent glasses involving mobile cations remain to be explored from the double-percolation perspective, as does relaxations in molecular models.

\section{Results}

\begin{figure*}[h!]
\begin{center}
        \includegraphics[width=14cm]{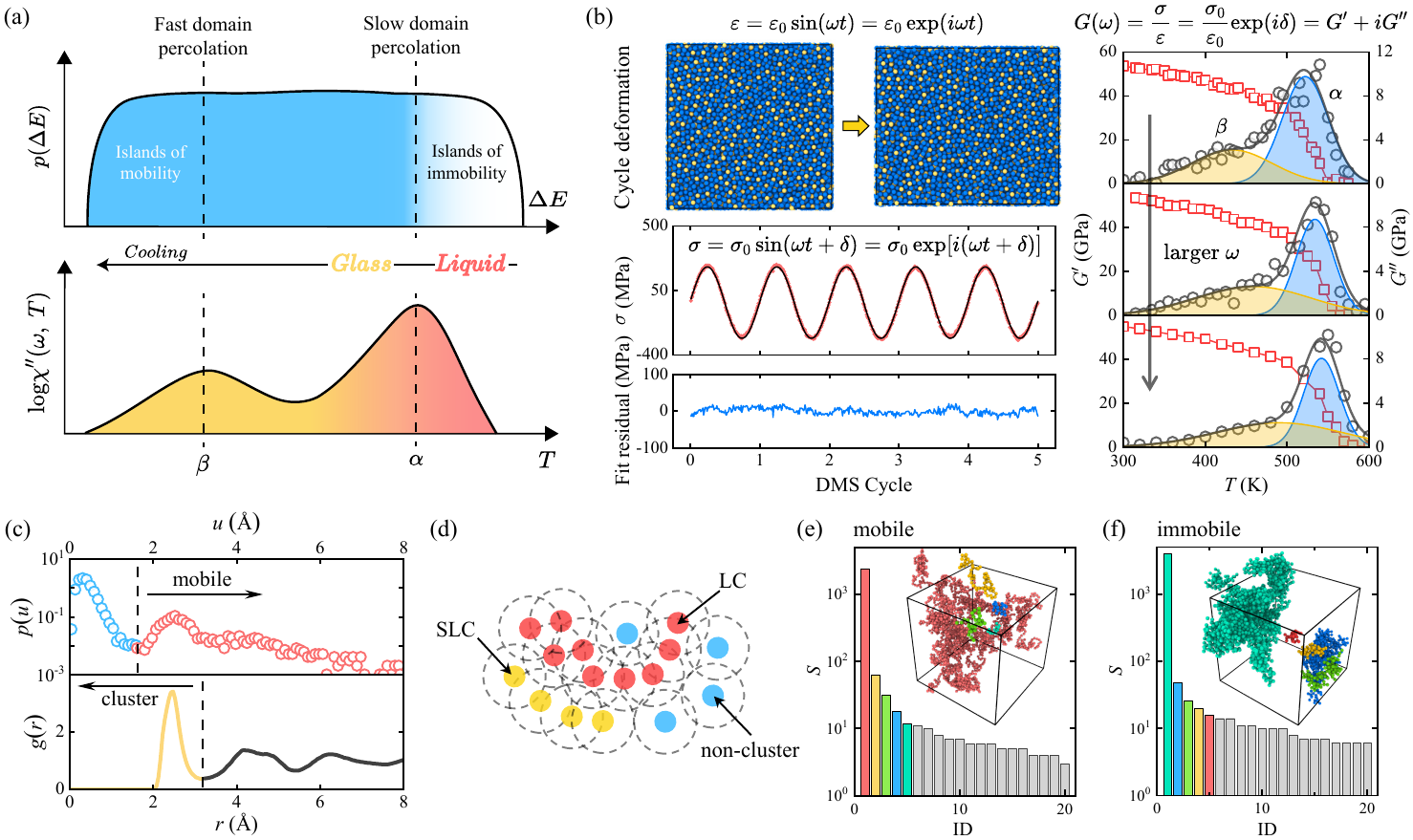}
        \caption{\label{fig:alfabeta}
        Main concepts. 
        (a) Generic double-percolation scenario for a glass-forming system's linear-response properties when cooled through the glass transition. The loss $\chi''(\omega)$ is probed at a fixed frequency during the cooling (lower panel), corresponding to activation energies that decrease with decreasing temperature (upper panel). Assuming that the activation-energy distribution is much wider than $k_BT$, at a certain temperature the high-activation-energy ``islands of immobility'' corresponding to times longer than the probe frequency will percolate. This locates the $\alpha$ process. At a lower temperature well within the glass phase, the low-activation-energy ``islands of mobility'' stop percolating, which pinpoints the $\beta$ relaxation \cite{ste10,gao23,dyr24}. 
        (b) Molecular Dynamics simulations mimicking experimental Dynamic Mechanical Spectroscopy for a $\textrm{Ni}_{80}\textrm{P}_{20}$ mixture. At selected times during the cooling, the sample is deformed periodically to determine the dynamic shear modulus $G(\omega)$. 
        (c) Definition of particles that are mobile on the time scale $\Delta t=2\pi/\omega$. The upper panel shows the distribution of all-particle displacements, the van Hove function $p(u,\Delta t)$. Particles with displacement larger than the minimum (dashed line) are designated as ``mobile'' and the remaining particles as ``immobile''. The lower panel shows the all-particle radial distribution function $g(r)$. Whenever two mobile/immobile particles are closer than the first minimum of $g(r)$, they are defined to belong to the same cluster \cite{hey89}. 
        (d) Example of the largest-particle cluster (LC, red) and second-largest-particle cluster (SLC, yellow), concepts that are defined for the mobile and the immobile particles, separately.
        (e) and (f) show examples of mobile and immobile particle clusters (insets) and their size ($S$) histograms.}
\end{center}
\end{figure*}

In DMS the glass transition is seen as a maximum in a plot of the temperature dependence of the fixed-frequency mechanical loss monitored during cooling, while the $\beta$ process manifests itself as a smaller peak below $T_g$. \fig{fig:alfabeta}(b) illustrates our simulation procedure. At selected times during the cooling, a binary 32000-particle sample is subjected to a small periodic elongation while the two transverse dimensions simultaneously are decreased to keep the volume fixed. This results in a periodic shear stress on a constant-volume sample, the magnitude and phase of which determine the complex frequency-dependent shear modulus $G(\omega)$ (more details are given in the Methods section and the Supplementary Information). The right side of panel (b) shows an example of $G(\omega)=G'(\omega)+iG''(\omega)$ at three frequencies, plotted as a function of temperature where squares and circles give the real and imaginary parts of $G(\omega)$. The $\beta$ process is most clearly visible at the lowest frequency (upper panel). 

There is a considerable freedom in how to define mobile and immobile particles. In order not to introduce arbitrary parameters we adopted the simple-minded approach of treating all particles on equal footing by proceeding as follows (\fig{fig:alfabeta}(c)). For a given time interval, $\Delta t$, mobile particles are defined via the all-particle van Hove function $p(u,\Delta t)$, which in all cases simulated has a well-defined first minimum or transition from peak to a long tail (Figs. S3 and S4). Particles with a displacement larger than this length are designated as ``mobile'', all other particles as ``immobile''. This depends on $\Delta t$, of course, which is put equal to $2\pi/\omega$ of the mechanical deformation. Note that in this approach any particle is either mobile or immobile. This is different from what is usually done; in Ref. \onlinecite{sta13}, for instance, less than one fifth of the particles were classified as either mobile or immobile.

Having defined mobile and immobile particles at any given time, two particles of same class are designated to belong to the same cluster if their distance is smaller than the minimum of the all-particle radial distribution function, $g(r)$, compare the lower panel of \fig{fig:alfabeta}(c). In this way, at any time during the cooling one identifies the largest cluster (LC) and the second-largest cluster (SLC) of mobile and immobile particles, respectively (\figs{fig:alfabeta}(d)-(f)). In summary, the probe frequency defines the time scale used to classify the particles as either mobile or immobile, and each of these two classes is subsequently divided into clusters.

\begin{figure*}[h!]
\begin{center}
        \includegraphics[width=13cm]{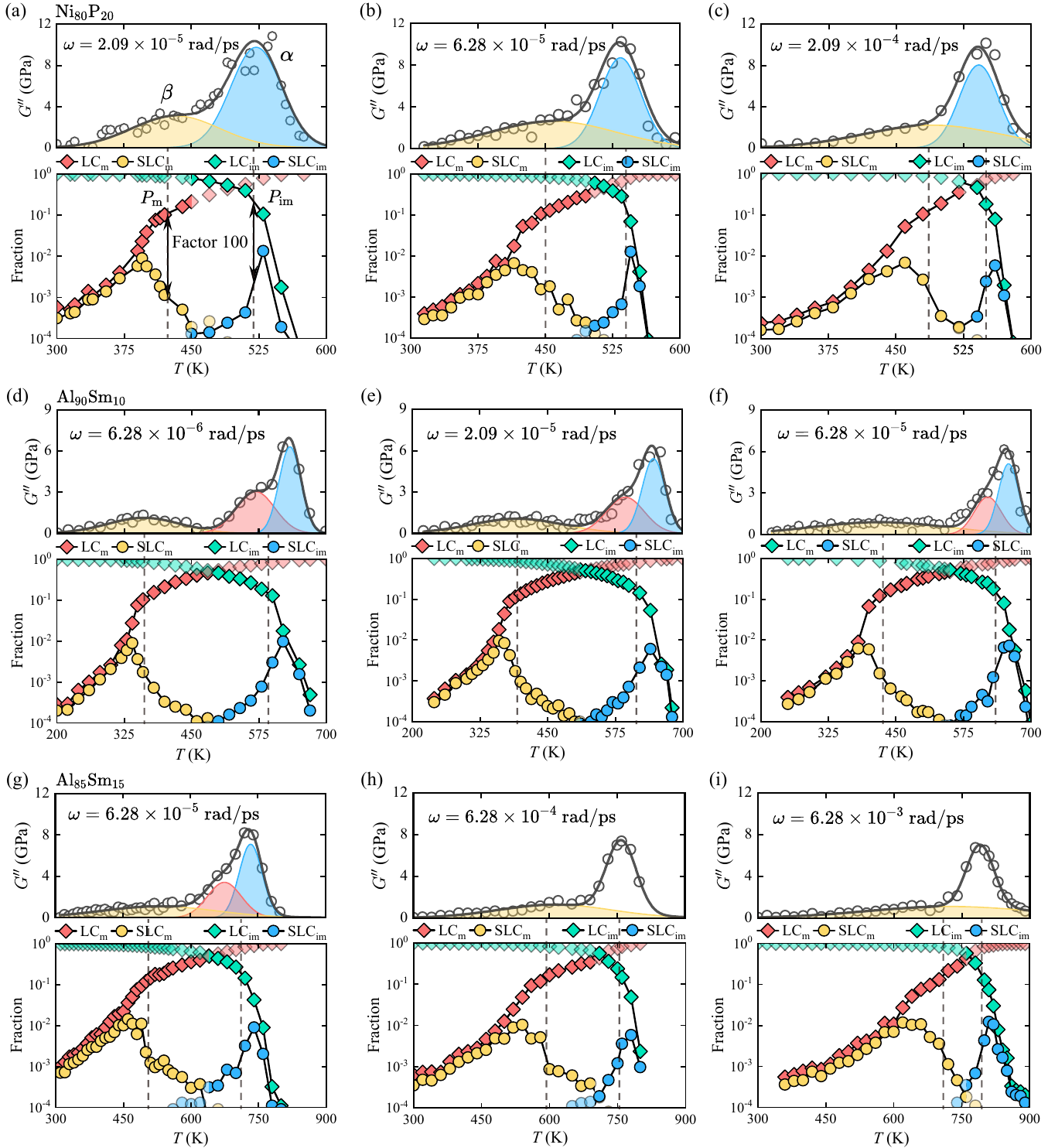}
        \caption{\label{fig:Ni80P20}
        Results at three frequencies for, respectively, $\textrm{Ni}_{80}\textrm{P}_{20}$ (a-c), $\textrm{Al}_{90}\textrm{Am}_{10}$ (d-f), and $\textrm{Al}_{85}\textrm{Sm}_{15}$ (g-i). For each subfigure the upper panel displays the shear-mechanical loss modulus $G''(\omega)$ at a fixed frequency monitored when cooling at 0.1 K/ns. The loss modulus is fitted to a sum of Gaussian peaks, corresponding to $\alpha$ (blue), $\alpha_2$ reflecting the $\alpha$ process asymmetry \cite{dyr24} (pink, not needed in all spectra), and $\beta$ (yellow) processes. The lower panels show the fraction of particles belonging to the largest cluster (LC) (pink/green) and the second-largest cluster (SLC) (yellow/blue) of mobile (left) and immobile (right) particles, respectively. The dashed lines mark the mobile- and immobile-particle percolation temperatures defined from the criterion that the LC is 100 times larger than the SLC.
        (h) and (i) do not show separate $\alpha$ and $\alpha_2$ processes because of the uncertainty of their relative positions (Fig. S5-S7 give more details on the $\alpha_2$ process and the fitting procedure).}
\end{center}
\end{figure*}

\Figs{fig:Ni80P20}(a), (b), and (c) show our results for $\textrm{Ni}_{80}\textrm{P}_{20}$ cooled through the glass transition and monitored at the three frequencies of \fig{fig:alfabeta}(b). Panel (a) gives results for the lowest frequency where the $\alpha$ and $\beta$ processes are best separated, while (b) and (c) give results for higher frequencies. The upper panels show the shear-mechanical loss modulus as a function of temperature during the cooling. The lower panels show the fraction of particles belonging to the $\textrm{LC}$ (orange and green diamonds) and $\textrm{SLC}$ (yellow and blue circles) of mobile and immobile particles, respectively (left and right). The two dashed vertical lines mark the mobile and immobile particle percolation temperatures, $T_{P_{\textrm{m}}}$ and $T_{P_{\textrm{im}}}$, defined by the criterion that the LC is 100 times larger than the SLC. 

Focusing first on the immobile particles, for all three frequencies we find that during cooling the LC and SLC fractions are virtually identical down to the temperature $T_{P_{\textrm{im}}}$ close to that of the $\alpha$ peak, below which LC completely dominates. Upon continued cooling into the glass phase an almost mirror behavior is found for the $\beta$ peak: Here the LC and SLC fractions are quite different down to a temperature, $T_{P_{\textrm{m}}}$, close to that of the $\beta$ peak, at which all the mobile-particle clusters become small ($\sim 1$\%). Thus the $\alpha$ peak is found where the immobile particles percolate and the $\beta$ peak is found where the mobile particles percolate.

\begin{figure*}[h!]
\begin{center}
        \includegraphics[width=13cm]{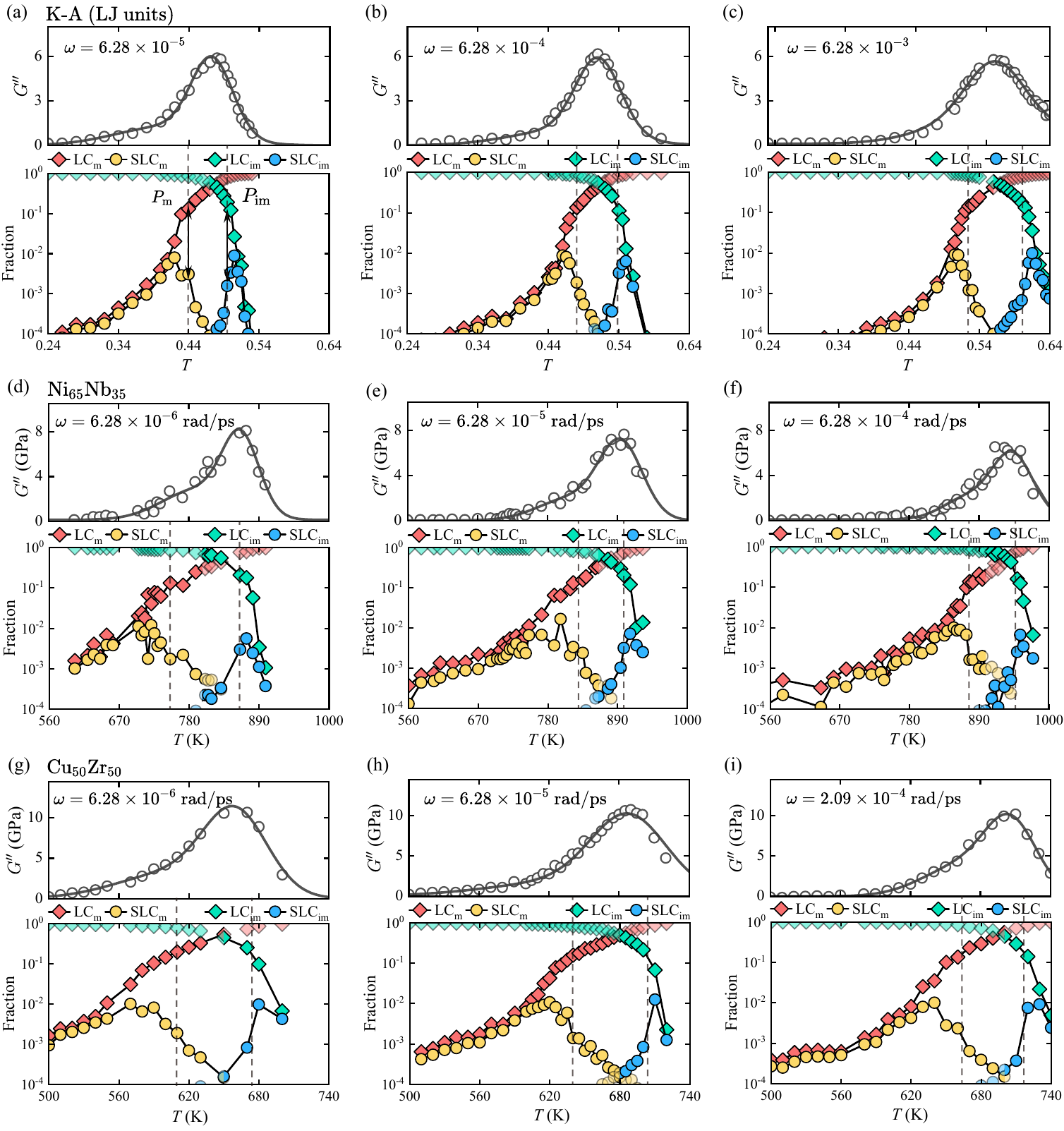}
        \caption{\label{fig:3dKA}
        Results at three frequencies for, respectively, 65:35 Kob-Andersen (K-A), $\textrm{Ni}_{65}\textrm{Nb}_{35}$, and $\textrm{Cu}_{50}\textrm{Zr}_{50}$ (similar data for the ternary system $\textrm{La}_{50}\textrm{Ni}_{35}\textrm{Al}_{15}$ are given in Fig. S9 of the Supplementary Information). The upper panels display $G''(\omega)$ for samples cooled at the rates 2$\cdot 10^{-7}$ (K-A, LJ units) and 0.1 K/ns ($\textrm{Ni}_{65}\textrm{Nb}_{35}$ and $\textrm{Cu}_{50}\textrm{Zr}_{50}$). As in \fig{fig:Ni80P20}, all $\alpha$ processes are found at the immobile-particle percolation temperature, but in contrast to \fig{fig:Ni80P20} these three samples have no well-defined $\beta$ process at the mobile-particle percolation threshold.}
\end{center}
\end{figure*}

To investigate the generality of these findings we carried out simulations of five other metallic glasses, $\textrm{Al}_{90}\textrm{Sm}_{10}$, $\textrm{Al}_{85}\textrm{Sm}_{15}$, $\textrm{Ni}_{65}\textrm{Nb}_{35}$, and $\textrm{Cu}_{50}\textrm{Zr}_{50}$,  $\textrm{La}_{50}\textrm{Ni}_{35}\textrm{Al}_{15}$, 
as well as of a Kob-Andersen-type binary Lennard-Jones mixture \cite{kob95} (\fig{fig:Ni80P20}, \fig{fig:3dKA}, and Fig. S9). The results can be summarized as follows:
1) The $\alpha$ process is in all cases characterized by immobile-particle percolation; 
2) Whenever there is a well-defined $\beta$ process in the form of a peak or a shoulder, it is characterized by mobile-particle percolation (\fig{fig:Ni80P20}); 
3) For the Kob-Andersen, $\textrm{Ni}_{65}\textrm{Nb}_{35}$, $\textrm{Cu}_{50}\textrm{Zr}_{50}$, and $\textrm{La}_{50}\textrm{Ni}_{35}\textrm{Al}_{15}$ systems, the $\beta$ process is not well separated from the $\alpha$ process and merely visible as a wing of the latter (\fig{fig:3dKA} and Fig. S9).

What is the difference between the systems with a clearly visible $\beta$ relaxation (\fig{fig:Ni80P20}) and those with only a wing (\fig{fig:3dKA})? Whenever the mobile- and immobile-particle percolation temperatures are close, one cannot expect to find well separated $\alpha$ and $\beta$ relaxations, and the $\beta$ process will be at most a wing of the $\alpha$ process. The percolation temperatures are reported in Table \ref{tab:T/T}; recall that these depend on the frequency/time scale in question. The table reveals a threshold of $T_{P_{\textrm{m}}}/T_{P_{\textrm{im}}}\cong 0.85$ below which the mobile- and immobile-particle percolation transitions are well enough separated for a $\beta$ relaxation to be identifiable, which is not the case above the 0.85 threshold. A genuine $\beta$ peak is seen for the systems with the lowest percolation-temperature ratios.

\Fig{fig:big}(a) illustrates our results by showing representative loss spectra in a plot where the $x$ coordinate is the ratio of the two percolation temperatures and the $y$ coordinate is the probe frequency. There is a pronounced $\beta$ process at low ratios of the percolation temperatures. When the two percolation temperatures are close ($T_{P_{\textrm{m}}}/T_{P_{\textrm{im}}} > 0.85$), on the other hand, the $\beta$ process is at most manifested as a wing. \Fig{fig:big}(b) summarizes our findings by reporting the percolation-temperature ratios for all simulations. 

We also simulated systems in 2D where one cannot at the same time have percolation of both the mobile and the immobile particles: If one type of particles percolate, their percolation cluster will necessarily sever any infinite cluster of the opposite type of particles (think of the paths and walls of a labyrinth – if the walls percolate, the paths do not, and \textit{vice versa}). Thus according to the double-percolation scenario, no separate $\alpha$ and $\beta$ processes should exist in 2D and the ratio of the two percolation temperatures should be close to unity. This is tested for five systems in \fig{fig:2d} and Figs. S16-S20 of the Supplementary Information (the 2D versions of the remaining two systems simulated in 3D, $\textrm{Ni}_{80}\textrm{P}_{20}$ and $\textrm{Ni}_{65}\textrm{Nb}_{35}$, could not be included in this analysis because they crystallized upon cooling). In no cases do we find a $\beta$ process, and the ratio of the two percolation temperatures is always significantly above 0.85 (actually it is slightly above unity, compare \fig{fig:big}(b)). 

\begin{figure*}[h!]
\begin{center}
        \includegraphics[width=14cm]{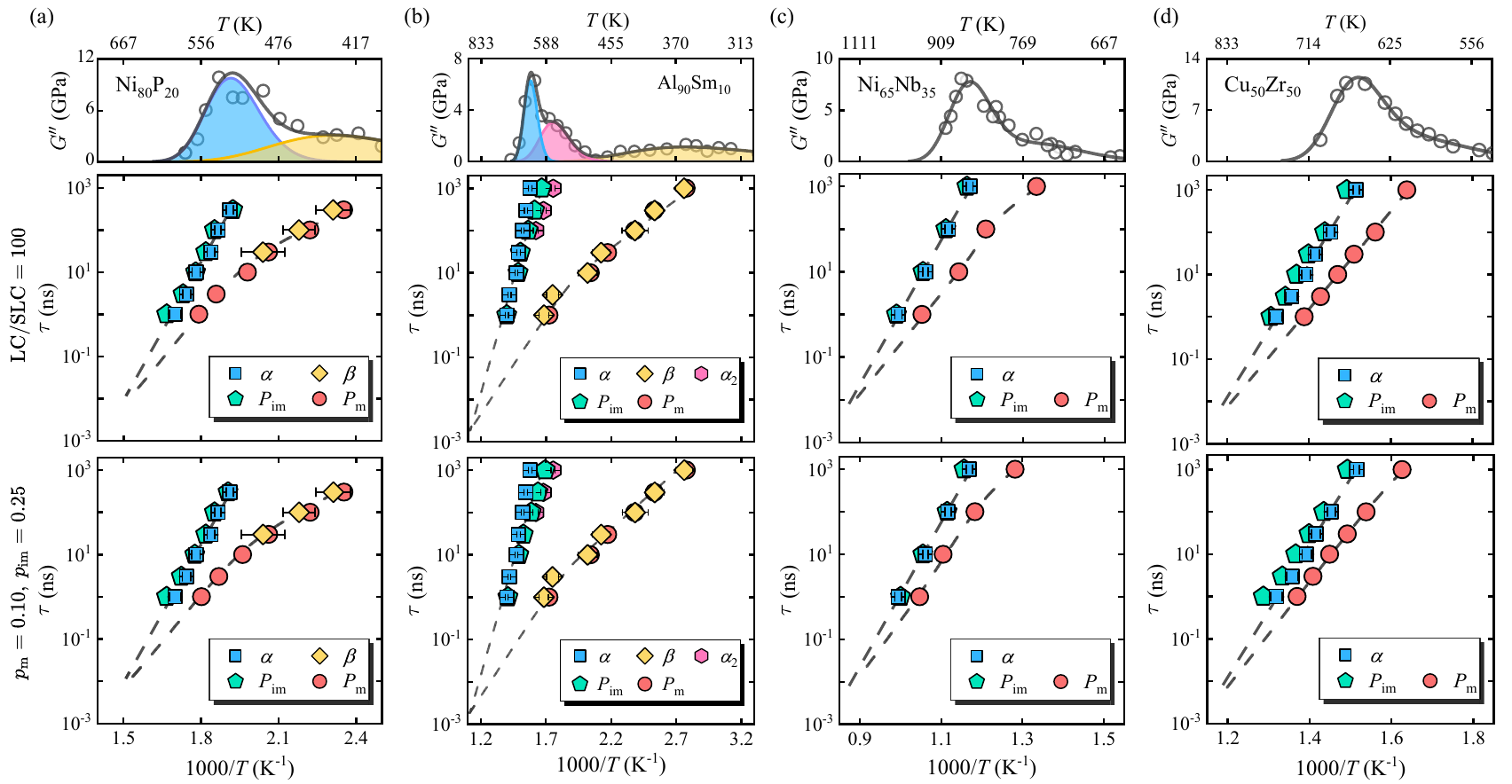}
        \caption{\label{fig:tau(T)}
        Temperature dependence of the $\alpha$ and $\beta$ average relaxation times and of the immobile- and mobile-particle percolation times. The figure shows data  for 
        (a) $\textrm{Ni}_{80}\textrm{P}_{20}$, 
        (b) $\textrm{Al}_{90}\textrm{Sm}_{10}$, 
        (c) $\textrm{Ni}_{65}\textrm{Nb}_{35}$, and
        (d) $\textrm{Cu}_{50}\textrm{Zr}_{50}$.
        The upper panels show examples of the loss moduli at a fixed frequency. For $\textrm{Ni}_{65}\textrm{Nb}_{35}$ and $\textrm{Cu}_{50}\textrm{Zr}_{50}$, the $\alpha$ and $\beta$ processes are not well separated, for $\textrm{Al}_{90}\textrm{Sm}_{10}$ there is an additional $\alpha_2$-process reflecting the asymmetry of the $\alpha$ process (pink hexagon in the lower panel). 
        The middle panels plot the percolation times defined as the inverse angular frequencies for which the temperature in question is the percolation temperature, as well as the relaxation times based on the  LC/SLC$\ = 100$ criterion used above. The $\alpha$ and $\alpha_2$ processes (blue square) give the temperature dependence of the average relaxation time (inverse loss-peak frequency), which is the same as that of immobile-particle percolation (green pentagon). Likewise, the $\beta$ process (yellow diamond) has the same temperature dependence of its average relaxation time as that of mobile-particle percolation (orange circle). The dashed lines are guides to the eye illustrating that the two processes merge at high temperatures. 
        The lower panels demonstrate the same picture if one instead uses the percolation thresholds derived from a detailed analysis of the percolation-cluster geometry (see Fig. S22 of the Supplementary Information): $P_{\textrm{m}}= 0.10$ and $P_{\textrm{im}}=0.25$ in which $P$ is the fraction of mobile/immobile particles. The error bars show the uncertainty of the fitting of peak positions for each process.}
\end{center}
\end{figure*}

Because the two percolation temperatures are defined by reference to the specific time scale $\Delta t=2\pi/\omega$, repeating the simulations for different frequencies allows one to identify the two time scales' temperature dependencies. Results are shown for four systems in \fig{fig:tau(T)} for frequencies covering more than 3 decades. The figure also plots the $\alpha$ and $\beta$ loss-maximum temperatures at the corresponding frequencies. Within the numerical uncertainty the latter coincide with the immobile- and mobile-particle percolation temperatures, respectively, confirming the connection between double percolation and mechanical response. The lower panels of \fig{fig:tau(T)} show that the same picture is seen if one uses an alternative percolation-temperature definition based on the observed percolation thresholds, 10\% and 25\% of the mobile and immobile particles, respectively (compare Fig. S22-S24 in the Supplementary Information; the different percolation thresholds reflect the interesting fact that the geometries of the immobile and the mobile percolation clusters differ).

\section{Discussion and outlook}

Percolation is important in many  contexts involving disordered solids \cite{zaccone_bog} by determining, e.g., thermodynamics, fragility, and stability of chalcogenide glasses \cite{phi85,tat90}, ac conduction at extreme disorder \cite{sch08}, spatial heterogeneity of soft modes \cite{sou09}, vibrational anomalies \cite{mar13b}, yielding \cite{gho17}, etc. This paper has investigated numerically a scenario according to which percolation also controls the two main relaxation processes of glass-forming liquids. The starting point is the assumption of extreme dynamic heterogeneity in the form of a wide barrier distribution for flow events (\fig{fig:alfabeta}(a)). From this one arrives at a picture characterized by percolation of the mobile and of the immobile particles. The former percolation transition is linked to the $\beta$ process and the latter to the main ($\alpha$) relaxation. Note that the proposed scenario does not take into account local facilitation, i.e., the recently discussed mechanisms according to which one flow event makes nearby flow events more likely by lowering their barriers, e.g., by long-ranged elastic interactions \cite{lem14,gui22,oza23,tah23,dyr24,has24}. 

To summarize our findings, extensive computer simulations of systems in three and two dimensions establish the following. Two temperatures can be identified marking the percolation of mobile and immobile particles, respectively, defined by reference to the particle displacements on a specific time scale. One percolation temperature marks the immobile-particle percolation threshold, which is found above the glass transition temperature; the second percolation temperature marks mobile-particle percolation taking place in the glass. Whenever the two percolation temperatures are well separated, they pinpoint the $\alpha$ and $\beta$ shear-modulus loss-peak temperatures for the frequency corresponding to the time scale in question. A ratio of approximately 0.85 of the two percolation temperatures separates two cases; whenever the ratio is below 0.85 the $\alpha$ and $\beta$ processes are well separated, whenever this ratio is above 0.85, the processes are partly or fully merged. In two dimensions, where one cannot have percolation of both mobile and immobile particles at the same time, there are no separate $\alpha$ and $\beta$ processes -- here the ratio of the two percolation temperatures is close to unity (in fact slightly above). 

While we have found no exceptions to this connection between percolation and the mechanical linear-response properties, it should be emphasized that one cannot yet conclude that a general, causal relation exists between the percolation of mobile and immobile particles and the $\alpha$ and $\beta$ processes. More work is needed before generality of the double-percolation picture can be concluded. It would be interesting to relate double-percolation to well-known features of the $\alpha$ and $\beta$ relaxations, in the hope that the scenario may help providing answers to questions like: Why is the $\beta$ relaxation Arrhenius while the $\alpha$ usually is not? How do the $\alpha$ and $\beta$ relaxations merge at high temperatures? Why is the $\beta$ process is usually symmetric in log(frequency)? -- Our conclusions are based on numerical studies of metallic glasses \cite{rut17}, confirmed by data on the Kob-Andersen model and the metalloid NiP. In order to investigate whether double-percolation always controls the $\alpha$ and $\beta$ processes of glass formers, for future work it will be important to simulate other inorganic/non-metallic systems, as well as more complex systems like molecular and polymeric glass formers.

\section*{acknowledgments}
The computational work was carried out on the public computing service platform provided by the Network and Computing Center of HUST. We thank for the support from the National Natural Science Foundation of China 52071147 (H.B.Y.). This work was also supported by the VILLUM Foundation’s \textit{Matter} grant VIL16515 (J.C.D.).

\section*{Author contributions}
H.B.Y. and J.C.D. devised the project. L.G. and H.B.Y. performed the simulations with input from T.B.S and J.C.D. All authors wrote and revised the manuscript.



\begin{table*}
\renewcommand{\arraystretch}{1.0}
\setlength{\tabcolsep}{8pt}
\centering
\caption{Data for all simulations showing that the $\beta$ process manifestation is predicted by the ratio between the mobile- and immobile-particle percolation temperatures. The $\beta$ process classification into ``Shoulder'', ``Wing'', or ``Peak'', is detailed in Fig. S8 of the Supplementary Information.}
\label{tab:T/T}
\begin{tabular}{cccccc}
\midrule
& $\boldsymbol{\omega}$ \textbf{(rad/ps)}  & \boldmath $T_{P_{\textbf{m}}}$ \textbf{(K)} & \boldmath $T_{P_{\textrm{im}}}$ \textbf{(K)} & \boldmath $T_{P_{\textbf{m}}}{/}T_{P_{\textrm{im}}}$  &  $\boldsymbol{\beta\,\,} \textbf{manifestation} $ \\ \midrule
\multicolumn{1}{c|}{$\textrm{Ni}_{80}\textrm{P}_{20}$}  & $2.09\times10^{-5}$ & 425 & 520  & {0.82} & {Shoulder}    \\
\multicolumn{1}{c|}{(EAM)}                                   & $6.28\times10^{-5}$ & 450 & 540  & {0.83} & {Shoulder}    \\
\multicolumn{1}{c|}{}                                   & $2.09\times10^{-4}$ & 485 & 550  & 0.88   & Wing   \\
\multicolumn{1}{c|}{}                                   & $6.28\times10^{-4}$ & 505 & 562  & 0.90   & Wing \\
\multicolumn{1}{c|}{}                                   & $2.09\times10^{-3}$ & 538 & 578  & 0.93   & Wing \\
\multicolumn{1}{c|}{}                                   & $6.28\times10^{-3}$ & 558 & 600  & 0.93   & Wing \\ \midrule
\multicolumn{1}{c|}{$\textrm{Al}_{90}\textrm{Sm}_{10}$} & $6.28\times10^{-6}$ & 365 & 595  & {0.61} & {Peak}        \\
\multicolumn{1}{c|}{(EAM)}                                   & $2.09\times10^{-5}$ & 395 & 615  & {0.64} & {Peak}        \\
\multicolumn{1}{c|}{}                                   & $6.28\times10^{-5}$ & 422 & 633  & {0.67} & {Peak}        \\
\multicolumn{1}{c|}{}                                   & $2.09\times10^{-4}$ & 460 & 660  & {0.70} & {Peak}        \\
\multicolumn{1}{c|}{}                                   & $6.28\times10^{-4}$ & 495 & 665  & {0.74} & {Peak}        \\
\multicolumn{1}{c|}{}                                   & $6.28\times10^{-3}$ & 585 & 708  & {0.83} & {Shoulder}    \\ \midrule
\multicolumn{1}{c|}{$\textrm{Al}_{85}\textrm{Sm}_{15}$} & $6.28\times10^{-5}$ & 505 & 710  & {0.71} & {Peak}        \\
\multicolumn{1}{c|}{(EAM)}                                   & $6.28\times10^{-4}$ & 590 & 755  & {0.79} & {Shoulder}    \\
\multicolumn{1}{c|}{}                                   & $6.28\times10^{-3}$ & 710 & 795  & 0.89   & Wing    \\ \midrule
\multicolumn{1}{c|}{$\textrm{Ni}_{65}\textrm{Nb}_{35}$} & $6.28\times10^{-6}$ & 750 & 860  & 0.87   & Wing    \\
\multicolumn{1}{c|}{(EAM)}                                  & $6.28\times10^{-5}$ & 827 & 900  & 0.92   & Wing \\
\multicolumn{1}{c|}{}                                   & $6.28\times10^{-4}$ & 875 & 948  & 0.92   & Wing \\
\multicolumn{1}{c|}{}                                   & $6.28\times10^{-3}$ & 950 & 1010 & 0.94   & Wing \\ \midrule
\multicolumn{1}{c|}{$\textrm{Cu}_{50}\textrm{Zr}_{50}$} & $6.28\times10^{-6}$ & 610 & 673  & 0.91   & Wing \\
\multicolumn{1}{c|}{(EAM)}                                   & $6.28\times10^{-5}$ & 640 & 703  & 0.91   & Wing \\
\multicolumn{1}{c|}{}                                   & $2.09\times10^{-4}$ & 662 & 717  & 0.92   & Wing \\
\multicolumn{1}{c|}{}                                   & $6.28\times10^{-4}$ & 680 & 735  & 0.93   & Wing \\
\multicolumn{1}{c|}{}                                   & $2.09\times10^{-3}$ & 700 & 755  & 0.93   & Wing \\
\multicolumn{1}{c|}{}                                   & $6.28\times10^{-3}$ & 720 & 780  & 0.92   & Wing \\ \midrule
\multicolumn{1}{c|}{$\textrm{La}_{50}\textrm{Ni}_{35}\textrm{Al}_{15}$}& $6.28\times10^{-5}$ & 527 & 590  & 0.89 & Wing \\
\multicolumn{1}{c|}{(DNN)}                                   & $6.28\times10^{-4}$ & 570 & 620  & 0.92 & Wing \\
\multicolumn{1}{c|}{}                                   & $6.28\times10^{-3}$ & 610 & 672  & 0.91 & Wing \\ \midrule
\multicolumn{1}{c|}{\textrm{K-A}}                       & $6.28\times10^{-6}$ & 0.40 & 0.46  & 0.87 & Wing \\
\multicolumn{1}{c|}{(LJ units)}                          & $6.28\times10^{-5}$ & 0.44 & 0.50  & 0.88 & Wing \\
\multicolumn{1}{c|}{}                                   & $6.28\times10^{-4}$ & 0.48 & 0.54  & 0.89 & Wing \\
\multicolumn{1}{c|}{}                                   & $6.28\times10^{-3}$ & 0.52 & 0.60  & 0.87 & Wing \\ \midrule
\end{tabular}
\end{table*}



\begin{figure*}[h!]
\begin{center}
        \includegraphics[width=14cm]{fig4.pdf}
        \caption{\label{fig:big}
         The ratio between mobile and immobile percolation temperatures determines the $\beta$ process manifestation. 
         (a) shows selected loss spectra of 3D mixtures at different frequencies. When the percolation-temperature ratio is below roughly 0.85 (marked in gray), the $\beta$ process is a ``Shoulder'' or a ``Peak'', while the $\beta$ process is absent or at most a ``Wing'' for percolation-temperature ratios larger than 0.85. The placement of each loss spectrum is determined by its maximum, as indicated by the two dashed lines.
         (b) summarizes all data for the percolation-temperature ratio and its correlation with the $\beta$ process manifestation. The data behind the box plots are shown nearby. Each box includes the middle 50\% of the data, i.e., the edges of each box mark the first and third quartile. The lines inside the boxes are the median and the whiskers mark the smallest and largest data values.}
\end{center}
\end{figure*}

\begin{figure*}[h!]
\begin{center}
        \includegraphics[width=17.5cm]{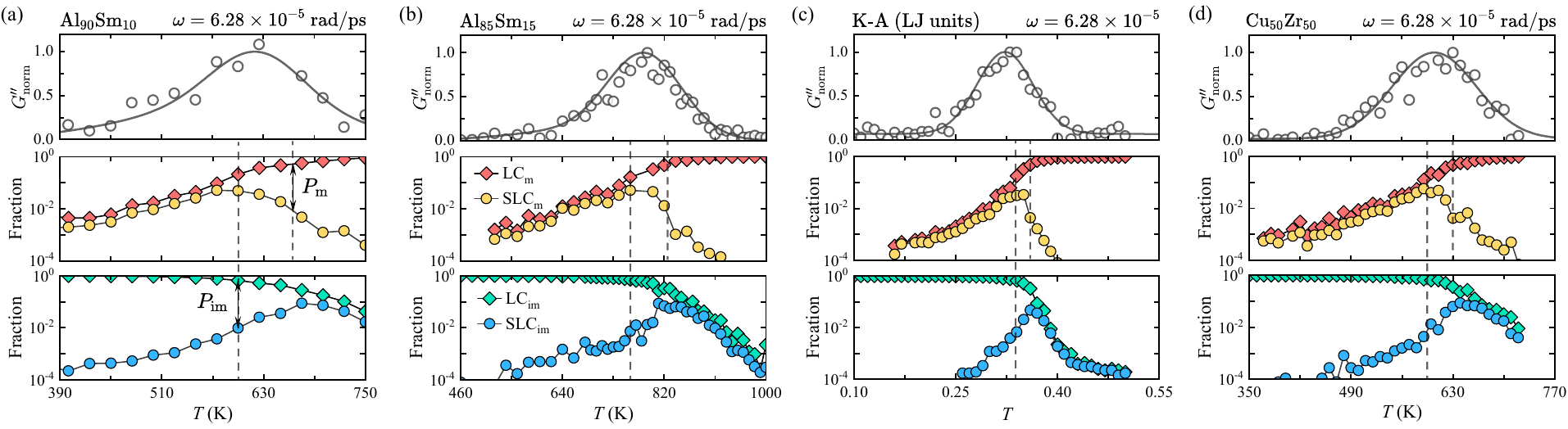}
        \caption{\label{fig:2d}
        Double-percolation studies of 2D models. 
        (a) gives data for $\textrm{Al}_{90}\textrm{Sm}_{10}$, 
        (b) for $\textrm{Al}_{85}\textrm{Sm}_{15}$, 
        (c) for the Kob-Andersen mixture, and 
        (d) for $\textrm{Cu}_{50}\textrm{Zr}_{50}$ (Fig. S20 gives data for the 2D $\textrm{La}_{50}\textrm{Ni}_{35}\textrm{Al}_{15}$ mixture).
        Contrary to in 3D, the two percolation temperatures are quite close, compare \fig{fig:big}(b). This is consistent with the finding that there are no separate $\alpha$ and $\beta$ processes (upper panels), reflecting the fact that in 2D mobile and immobile-particle percolation cannot occur at the same time.}
\end{center}
\end{figure*}

\clearpage

\clearpage 

\section*{Methods}
\noindent{\textbf{Molecular Dynamics simulations and dynamic mechanical spectroscopy}}

Twelve different mixtures were simulated using LAMMPS \cite{lmp22}. The 3D mixtures are: 
$\textrm{Ni}_{80}\textrm{P}_{20}$ (32000 particles), 
$\textrm{Al}_{90}\textrm{Sm}_{10}$ (32000/10976/2916 particles), 
$\textrm{Al}_{85}\textrm{Sm}_{15}$ (9088 particles), 
$\textrm{Ni}_{65}\textrm{Nb}_{35}$ (9088 particles), 
$\textrm{Cu}_{50}\textrm{Zr}_{50}$ (32000 particles), 
$\textrm{La}_{50}\textrm{Ni}_{35}\textrm{Al}_{15}$ (8788 particles), 
and 65:35 Kob-Andersen (70304 particles).
The 2D mixtures are: 
$\textrm{Al}_{90}\textrm{Sm}_{10}$ (7200 particles), 
$\textrm{Al}_{85}\textrm{Sm}_{15}$ (7200 particles), 
$\textrm{Cu}_{50}\textrm{Zr}_{50}$ (7200 particles), 
$\textrm{La}_{50}\textrm{Ni}_{35}\textrm{Al}_{15}$ (7200 particles), 
and 65:35 Kob-Andersen (7200 particles).
It is demonstrated by example in Fig. S21 of the Supplementary Information that these samples are large enough to represent the genuine bulk response \cite{sch24}, with the correct percolation temperatures. With the exception of the Kob-Andersen and $\textrm{La}_{50}\textrm{Ni}_{35}\textrm{Al}_{15}$ mixtures, which use Lennard-Jones and Deep Neural Network (DNN) potentials \cite{Zhang24}, respectively, all mixtures employ embedded-atom-method (EAM) potentials \cite{lmp13}. 

Simulations were initiated using the melt-quench method that involves the following two steps: 1) annealing the mixture at a high temperature above the melting point until its energy stabilizes, thereby producing a high-temperature equilibrium liquid; 2) cooling this liquid to below room temperature (the final temperature of the Kob-Andersen mixtures is below 0.2). The time step of the annealing is 1 fs while it is 2 fs for the cooling (0.001 and 0.002 for Kob-Andersen mixtures, respectively). The simulations employed periodic boundary conditions, used a Nose-Hoover thermostat, and most were performed at constant pressure and temperature; the 2D mixtures and the 3D Kob-Anderson mixture were simulated at constant volume and temperature, however.

The simulations were designed to numerically mimic the protocol of real DMS experiments. Thus a sinusoidal volume-preserving strain $\varepsilon(t) = \varepsilon_0 \textrm{sin}(\omega t)$ is applied at selected times, with strain amplitude $\varepsilon$ along the $x$ or $xy$ direction of the simulation box (leading to the same results). The resulting shear stress $\sigma$ is fitted by $\sigma(t) = \sigma_0 \textrm{sin}(\omega t + \delta)$. The storage and loss moduli are calculated from $G' = \sigma_0 / \varepsilon_0 \textrm{cos} (\delta)$ and $G'' = \sigma_0 / \varepsilon_0 \textrm{sin} (\delta)$, respectively. The deformation amplitude must be small enough to be within the linear-response regime. To ensure this we used a deformation of 1.4\%; the two transverse dimensions were simultaneously decreased by 0.7\% to maintain the volume. To strike a balance between system stability and simulation duration, a time step of 10 fs is utilized when probing the shear-mechanical properties. Table S1 of the Supplementary Information summaries the details of the molecular dynamics simulations.

\noindent{\textbf{Cluster and percolation analysis}}

Our percolation analysis involves \textit{all} particles by dividing them into two classes, mobile and immobile. The division is controlled by the magnitude of the displacement, $u$, of each particle over one molecular dynamics DMS cycle. In order to identify a suitable critical displacement to distinguish between mobile and immobile particles, $u_c$, the van Hove function $p(u) \equiv [P(u + \Delta u) - P(u)]/\Delta u$ is used in which $P(u)$ is the cumulative distribution giving the probability of finding the value $X \leq u$, normalized according to $\int_0^{\infty} p(u)du=P(\infty)=1$. The threshold displacement $u_c$ is defined from the first minimum (or transition to long tail) of $p(u)$; by definition all mobile particles have a displacement greater than $u_c$ and all immobile particles a displacement smaller than $u_c$. A cluster consists of particles ``close'' to neighbors of the same class (mobile or immobile), defined by reference to the all-particle radial distribution function $g(r)$: When the distance between two particles is below that of the first minimum of $g(r)$, $r_c$, the particles by definition belong to the same cluster. The values of $u_c$ and $r_c$ depend on the frequency, thermodynamic state point, and system in question, but for all systems these values are almost identical at different temperatures and frequencies (compare Fig. S4 of the Supplementary Information). Table S2 reports $u_c$ and $r_c$ for all mixtures.

\section*{Data availability}
All data behind the figures are available at \url{https://doi.org/10.5281/zenodo.13925660}.

\section*{Code availability}
The simulation package LAMMPS (\url{https://www.lammps.org}) was used for all molecular dynamics simulations. The code and scripts of the analysis are available at \url{https://doi.org/10.5281/zenodo.13925660}.

\end{document}